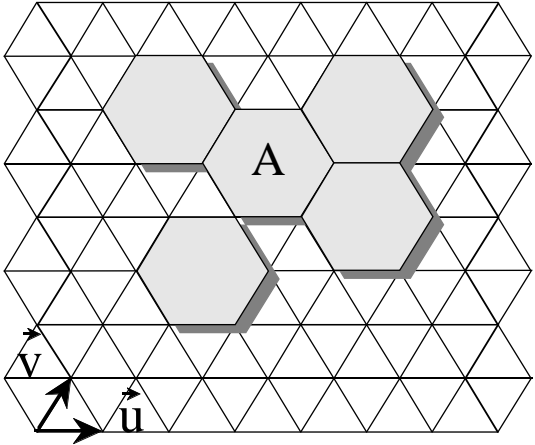
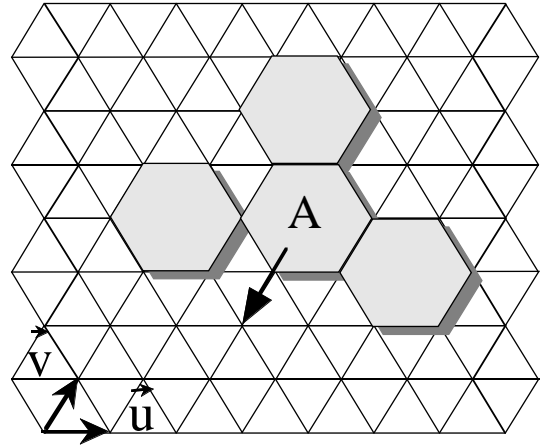

Fig. 1a        Fig. 1b

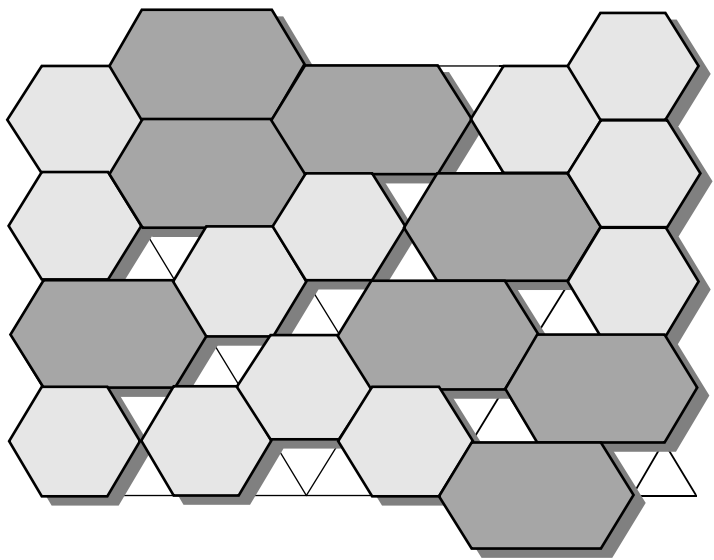

Fig. 2

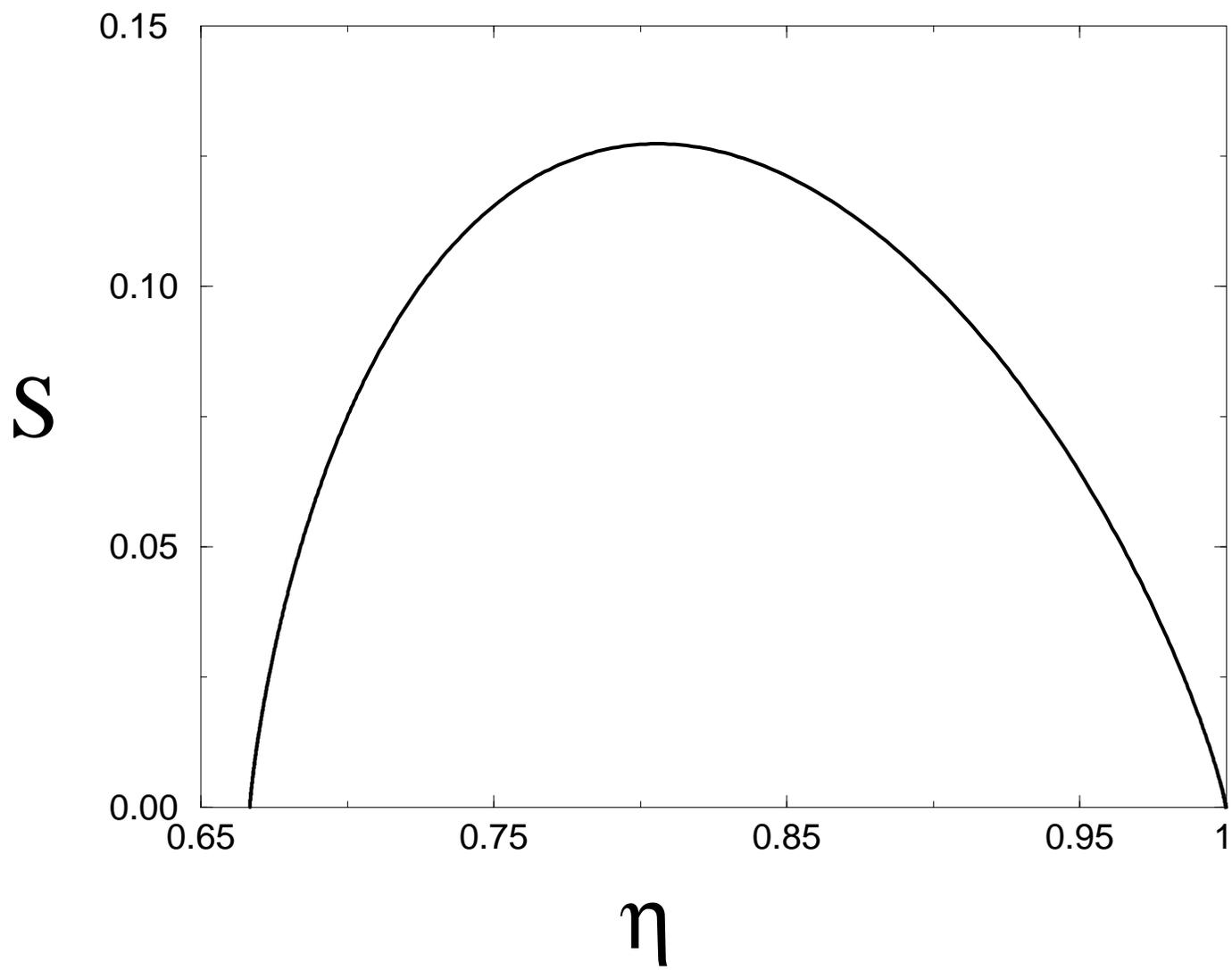

Fig. 3

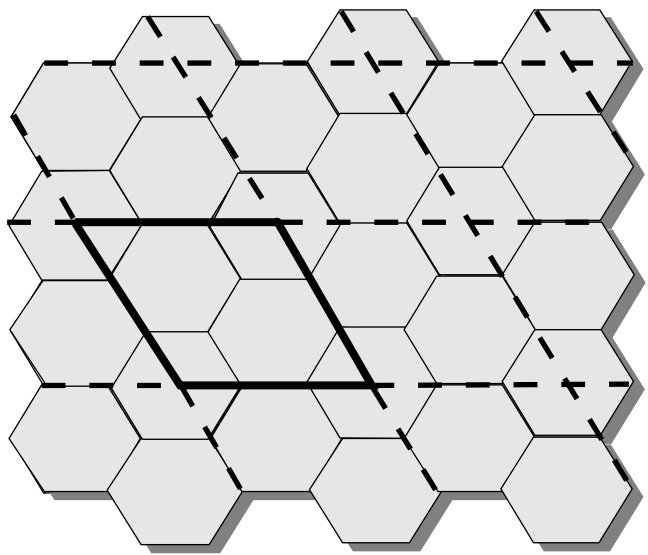

Fig. 4a

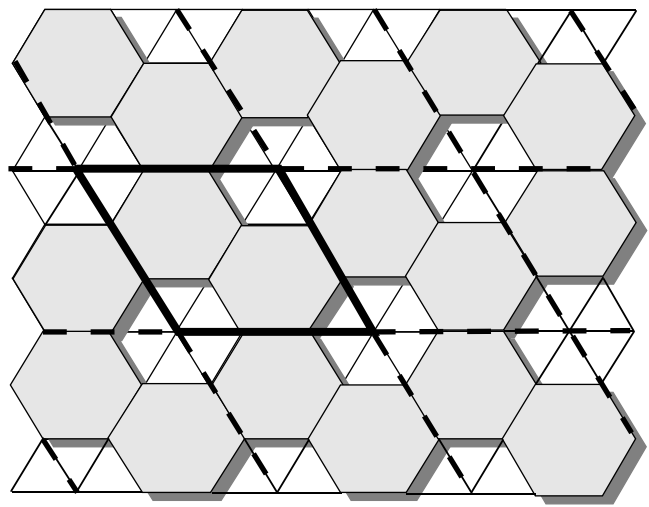

Fig. 4b

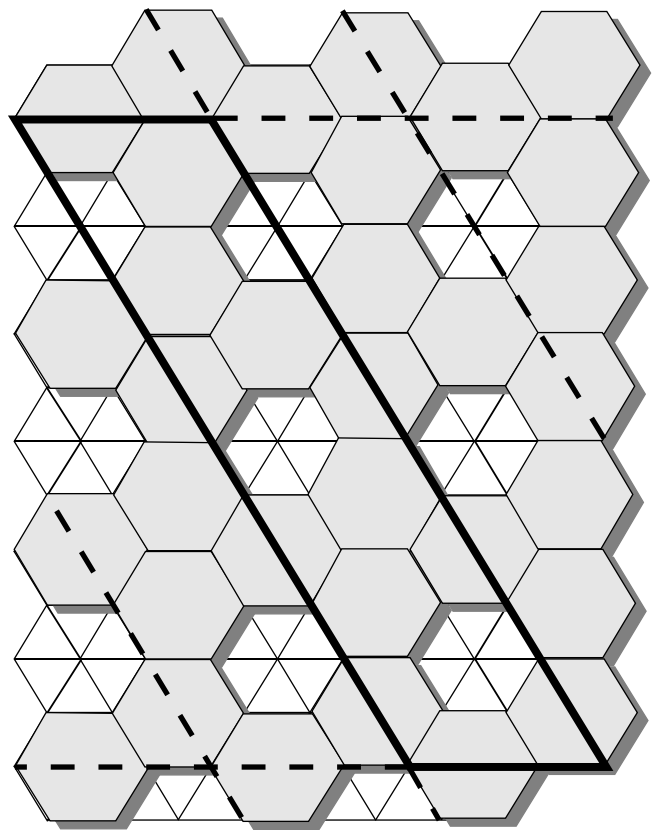

Fig. 4c

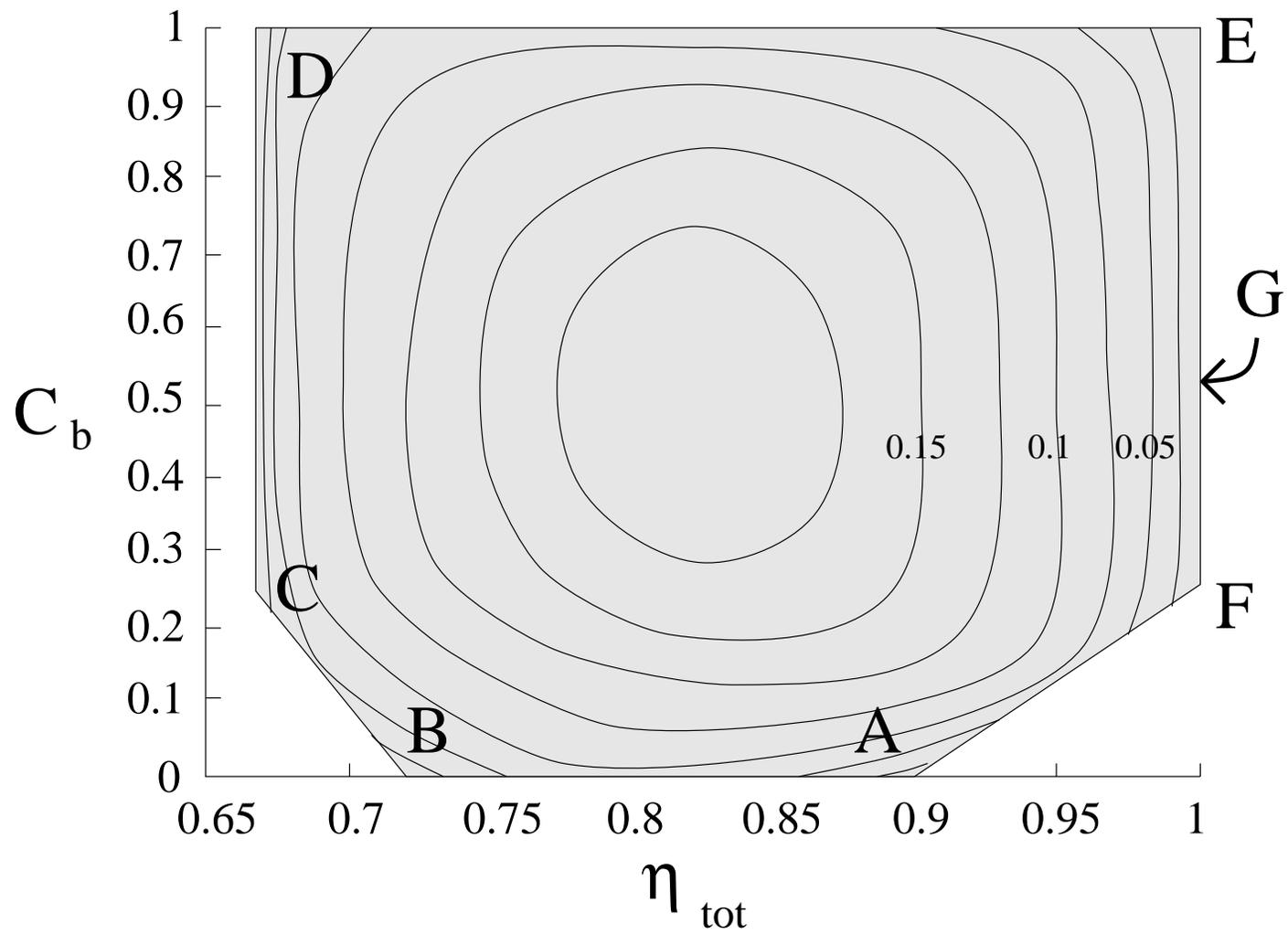

Fig. 5

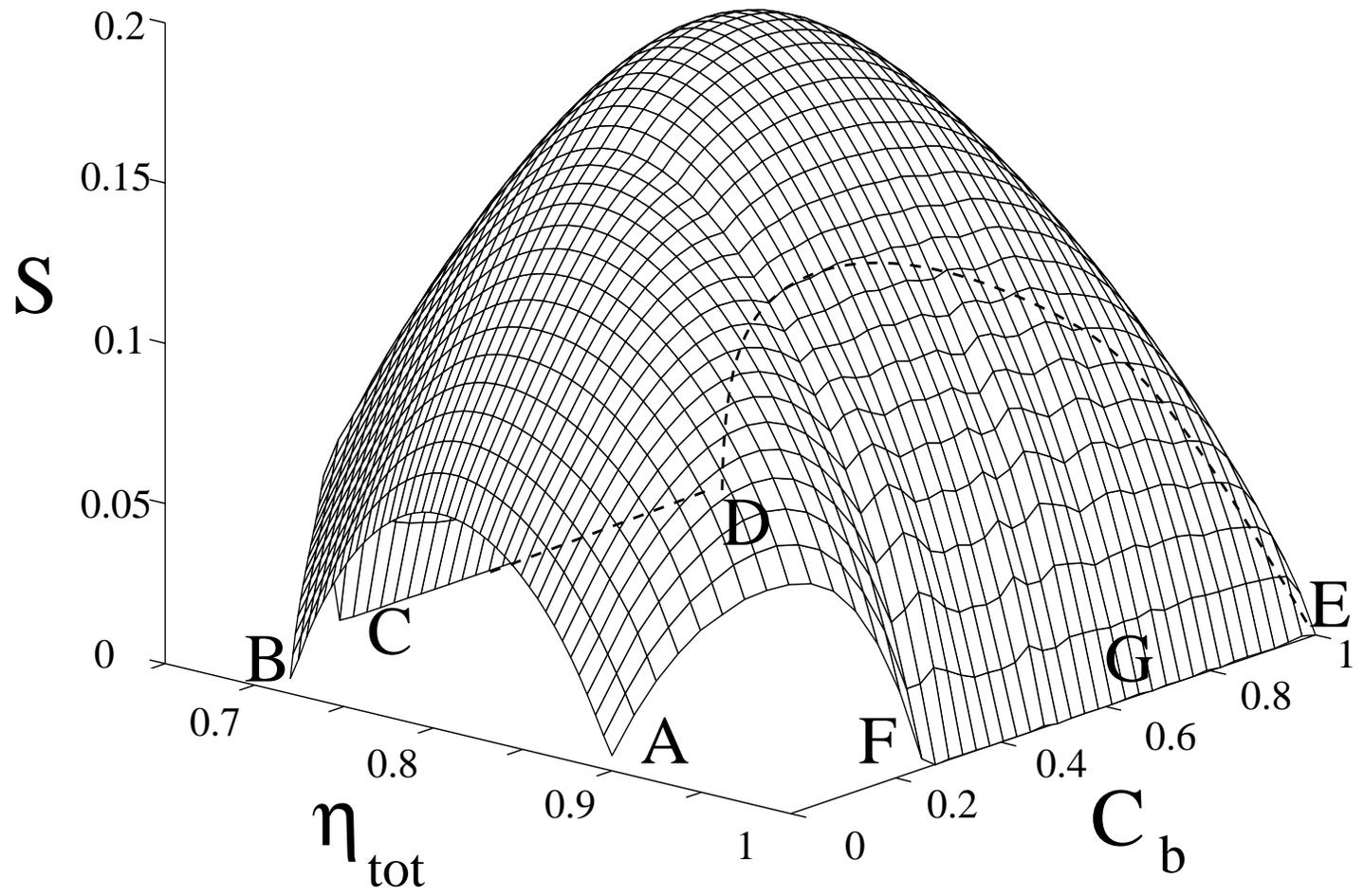

Fig. 6

# Entropy of particle packings : an illustration on a toy model.

Rémi Monasson* and Olivier Pouliquen†

* CNRS - Laboratoire de Physique Théorique de l'ENS, 24 rue Lhomond, 75231 Paris Cedex 05, France

† Laboratoire d'Hydrodynamique LadHyX, Ecole Polytechnique, F-91128 Palaiseau Cedex, France

(July 31, 1996)

## Abstract

A toy model of particles packings is presented, which consists in arranging hexagons on a triangular lattice according to local stability rules. The number of stable packings is analytically computed and found to grow exponentially with the size of the lattice, which illustrates the concept of packing entropy first proposed by Edwards and collaborators. The analysis is carried out for both the monodispersed case and the more interesting, i.e. more disordered, bidispersed case.

# I. INTRODUCTION

Particle packings are present in many industrial problems involving powders or granular materials [1], and have also been extensively studied in physics since hard sphere packings have been proposed as a model of simple liquids [2]. One of the major difficulties arising in the description of packings is that, for a given macroscopic external stress such as gravity or a confining pressure, there exist a multiplicity of arrangements, all stable from the mechanical point of view. As for granular media thermal effects are negligible, these metastable states cannot decay down to lower energy states and are therefore observable, depending on the dynamics used to create the packings.

For example in the extensively studied problem of the regular spheres packings under gravity [3], arrangements can be build with different volume fractions. Gently rolling the particles in a container gives rise to the loose packing with a volume fraction around 0.6 [3,4] (which goes down to 0.55 in the limit of zero gravity [5]), whereas vibrating the particles yields the random close packing [2–4,6–14] with a volume fraction of 0.635, which is the highest density that can be reached by collectively handling the particles. Higher volume fractions can be obtained by individually deposing the spheres in an ordered fashion, the maximum being attained for the faced–centered cubic lattice [15,16]. In order to acquire a better understanding of particles packings it is thus necessary to first determin and characterize the numerous packings that are mechanically stable, before studying the role of the dynamics in the selection of a particular packing.

Counting and characterizing metastable states is a problem often encountered in statistical physics of disordered systems, e.g. spin glasses or neural networks. Such systems are indeed known to display very complicated free–energy landscapes leading to a huge number of metastable states, in which they can get trapped depending on the particular dynamics considered. Due to the striking analogy with the particles packings problem, it is thus tempting to approach the latter in a statistical mechanics framework. The first attempt has been proposed by Edwards and collaborators [17–20]. Their approach is based on an analogy between the packing volume $V$ and the energy $E$ in conventional statistical systems. They introduced the concept of the packing entropy $S$ as the logarithm of the number of packings at a given volume and defined the compactivity $X = \partial V/\partial S$, that is the corresponding (microcanonical) "temperature". These attractive concepts have been applied to the study of packing of spheres or packing of mixture of particles. Unfortunately the calculations are not straightforward and have been carried out so far only at a cost of drastic simplifications [20]. Another difficulty arising from this analogy between particles packings and classical thermodynamics systems comes from the notion of compactivity which plays the role of the temperature. Whereas the temperature is a parameter that can easily be controlled in conventional thermodynamics system, the compactivity does not seem to be an accessible parameter in granular packing. This discrepancy leads to some difficulties in the interpretation of the results found within this approach.

In this paper we are interested in the study of the packing statistics in a two dimensional lattice model. By opposition with the studies cited above, our model is simple enough to allow an exact calculation of the statistics of the stable configurations without any a priori analogy with usual statistical mechanics. It therefore represents a good toy model to discuss the relevance of some concepts such as the entropy or the compactivity for particles packings. Our model is a modified version of the hard hexagon model first introduced to study the liquid–solid transition and analytically solved by Baxter fifteen years ago [21]. It consists in arranging hexagons on a triangular lattice according to some realistic local stability rules. The whole statistics of the stable packings can be calculated using a transfer matrix method [22] to estimate the number of arrangements at a given density of occupation. We present results for both a monodispersed medium, and a bidispersed medium. In the monodispersed case the boundary conditions are found to play a major role and no thermodynamics limit can be found. This feature disappears in the bidispersed case and we show that in the limit of large lattice the entropy is indeed extensive for volume fractions lying between a lower and upper bounds, which coincide with the loose and close packings respectively.

The hexagon model and the packing entropy are defined in Section II, while the method of calculation is exposed in Section III. Results for monodispersed packings of hexagons and for a binary mixture are presented respectively in section IV and V. In Section VI we present some conclusive remarks.

## II. THE HEXAGON PACKING MODEL

The classical hard hexagon model was first introduced as a model of simple liquid. It consists in hexagons placed at the nodes of a triangular lattice (Fig.1). The interactions are of hard–core type and prevent the hexagons from overlapping each other. Remarkably, the properties of this two–dimensional model may be analytically studied in the limit of infinite lattices [21]. In particular, the hard hexagon system has been shown to exhibit a liquid–solid transition [21–23] at high densities.

In this paper, we introduce a version of the hexagon model suitable for studying the properties of particles packings. An essential difference between the liquid and the packing problematics is that, in the latter case, there must exist a contact network between the particles in order to resist external stresses (e.g. gravity, confining pressure, ...) and achieve the mechanical equilibrium. Our procedure is thus to introduce a stability condition in the hard hexagon model exposed above, and to discard the configurations which we shall judge as *unstable* configurations and to keep only the remaining *stable* ones. The exact stability condition would be to calculate all the forces between the particles, and to check the mechanical stability of each particle. However, the calculation of the forces network in particles packing is still an open problem [24–27] far beyond the scope of this paper. Moreover, our purpose is to develop a model simple enough to allow theoretical calculations. We therefore choose to apply a local geometrical stability rule instead of the real stability criterion. The rule adopted in our model is presented in Fig.1 : an hexagon is considered to be stable if blocked by other hexagons, that is if all six local moves along the network axis are forbidden. In Fig.1a, hexagon A is considered as stable whereas in Fig. 1b it is unstable since it can be moved along the arrow. A stable configuration is a packing for which all the hexagons are stable in the previous sense. The same kind of realistic but ad hoc stability conditions have already been adopted in other studies [28,29].

As we stressed in the introduction, we shall mainly be interested in counting the number of packings as a function of their densities. In the following, we call density of a packing $\rho$ the number of hexagons it contains normalized with respect to the total number $N$ of lattice sites. A more natural quantity is the packing fraction $\eta$ which characterizes the fraction of the total area of the lattice occupied by the hexagons of the packing. The packing fraction is simply related to the density through $\eta = 3\rho$. Note that the proportionality factor three entering the previous expression only depends on the geometry of the particles (and has to be modified in the case of polydispersed packings - see below). We then define $C(\rho, N)$ the number of packings with density equal to $\rho$ that satisfy both hard–core repulsion and stability conditions . As usual in statistical mechanics, $C(\rho, N)$ is a combinatorial quantity that is expected to scale exponentially with the number of sites $N$ of the lattice. We therefore define the entropy $S$ at a density $\rho$ as the (normalized) logarithm of the number of possible packings having this density

$$S(\rho) = \lim_{N \to \infty} \frac{1}{N} \ln C(\rho, N) \quad . \tag{1}$$

Note that this definition is less straightforward that it might appear at first sight. The existence of a well–defined thermodynamic limit when $N \to \infty$ is not assured at all. We shall come back to this important point in Section IV.

Up to now, our packings include identical hexagons only. A natural extension of the model is to consider polydispersed packings, introducing a new type of particle drawn Fig.2 (for simplicity, the large side of the latter will always be parallel to the horizontal axis of the lattice). In the following, we shall call big particles (or big hexagons) the new particles, while small particles (or small hexagons) will refer to the hexagons considered so far. The stability criterion for the big hexagons remains unchanged, namely all local moves are forbidden. The packing is now characterized by the two densities $\rho_s$ and $\rho_b$ respectively defined as the numbers of small and big particles per site. The corresponding packing fraction of big particles is obviously given by $\eta_b = 5\rho_b$. As previously, we may define the entropy $S(\rho_s, \rho_b)$ as the (normalized) logarithm of stable packings at fixed densities of small and big particles.

## III. ANALYTICAL METHOD

The entropy (1) being a microcanonical quantity, its calculation is a difficult task. A possible way to circumvent this difficulty is to introduce the generating function

$$\Xi(z, N) = \sum_{n=1}^{N} z^n \, C\left(\frac{n}{N}, N\right) \tag{2}$$

where the new variable $z$ is called fugacity. For small fugacities $z \ll 1$, the sum (2) is dominated by low $n$, i.e. sparse packings while large $z$ favourish high density stable configurations. Therefore, in the thermodynamical limit, the knowledge of the function

$$p(z) = \lim_{N \to \infty} \frac{1}{N} \ln \Xi(z, N) \tag{3}$$

for all fugacities $z > 0$ allows to obtain a parametric representation $\{\rho(z), S(z)\}$ of the entropy curve $S(\rho)$ through the Legendre identities

$$\begin{aligned} \rho(z) &= z \, \frac{dp}{dz}(z) \\ S(z) &= p(z) - z \, \frac{dp}{dz}(z) \, \ln z \end{aligned} \tag{4}$$

One may notice that the slope of the curve at fugacity $z$

$$\left. \frac{dS}{d\rho} \right|_{\rho(z)} = -\ln z \tag{5}$$

is related to the compactivity $X$ introduced by Edwards and his collaborators [17–20] through the relation $X = 1/\ln z$. Contrary to the previous authors, we however stress that the entire curve $S(\rho)$ is spanned when $z$ runs from zero to infinity and that no a priori restriction on the sign of the slope (5) has to be imposed.

In order to calculate the generating function (2), we take advantage of the locality of the interactions between the particles and use a transfer matrix method. To do so, we impose periodic boundary conditions to the lattice. It thus lies on a parallelogram comprised of $m$ columns and $L$ ranks (with $N = m \times L$), see Fig.4. The transfer matrix method has been used in the hard hexagon liquid case and is exposed in [22]. Briefly speaking, it consists in propagating the hard–core repulsion from lines to lines, which can be written in a matricial way. We have modified this approach to take into account the stability criterion in addition to the non overlapping condition. The structure of the matrix is more complicated due to the next nearest neighbor interactions. In this framework, the generating function reads $\Xi(z, m, L) = \text{Trace } [\mathcal{Q}_m(z)^L]$, where the transfer matrix $\mathcal{Q}_m$ depends on the fugacity $z$. Therefore, once we have computed $\mathcal{Q}_m(z)$, the limit $L \to \infty$ is straightforward since the problem reduces to finding the largest eigenvalue $\Lambda(z, m)$ of $\mathcal{Q}_m(z)$. If the latter is non degenerate, it must be real and we obtain

$$\Xi(z, m, L) \underset{L \gg 1}{\simeq} \Lambda(z, m)^L \qquad \text{(non degenerate real case)} \qquad (6)$$

However, since $\mathcal{Q}_m$ contains zero elements, the largest eigenvalues (in modulus) may be complex. In this case, there are $q$ eigenvalues having the same (maximal) modulus $\Lambda(z, m)$ but different arguments $\theta_1(z, m), \ldots, \theta_q(z, m)$. The generating fonction then reads

$$\Xi(z, m, L) \underset{L \gg 1}{\simeq} \Lambda(z, m)^L \sum_{j=1}^{q} e^{i\theta_j(z,m)} \qquad \text{( q–fold complex case)} \qquad (7)$$

In both cases, we deduce the entropy curve of the stable packings for a lattice of width $m$ and infinite length from relations (4), (6) and (7). All the results which will be exposed in this paper thus correspond to packings lying on an infinite (tilted) vertical cylinder ($L \to \infty$) including $m$ sites on its perimeter. In practice, $\Lambda(z, m)$ is obtained by diagonalizing the transfer matrix numerically, or analytically for small $m$. The size of $\mathcal{Q}_m$ growing exponentially with $m$, the largest widths we could reach extend to $m \simeq 10$ typically. Fortunately, this is already sufficient to draw some conclusions as shown in next section.

In the case of polydispersed packings, the calculation of the entropy $S(\rho_s, \rho_b)$ requires the introduction of a generating function $\Xi(z_s, z_b, N)$ similar to (2) with an additional fugacity $z_b$ accounting for big particles. The use of two fugacities is necessary to compute the two densities and thus obtain the complete description of the bidispersed packings. As a consequence, the studies using a single compactivity [17–20] are able to determin the total density only, the relative concentrations of the species being uncontrolled.

## IV. RESULTS FOR MONODISPERSED PACKINGS

The calculation of the entropy can be analytically carried out for narrow systems, namely when the width of the lattice $m$ is small. First, we present the results obtained for $m = 3$. The system is in this case too small to be representative of the thermodynamics limit we are interested in. However, the analytical calculation of the entropy for $m = 3$ leads to non trivial results on the statistics of packing, and moreover brings to light some problems concerning the thermodynamics limit, problems that are related to the periodicity of the stable packings made of monodispersed particles.

The non zero eigenvalues $\lambda$ of $\mathcal{Q}_3(z)$ are the roots of the polynomial

$$P(\lambda, z) = \lambda^9 - (3z^2 + z^3)\lambda^6 + 3z^4\lambda^3 - z^6 \quad . \tag{8}$$

As a consequence, if $\lambda$ is an eigenvalue, so are $\lambda\, e^{2i\pi/3}$ and $\lambda\, e^{-2i\pi/3}$. This degeneracy of the modulus of the eigenvalues implies that the number $L$ of sites of the lattice in the vertical direction has to be a multiple of three in order to get a non zero number of stable configurations. This periodicity problem will be discussed later in the paper.

The modulus of the largest three eigenvalues is

$$\Lambda(z) = \frac{z}{3}\left[1 + \left(1 + \frac{27}{2z}\left(1 + \sqrt{1 + \frac{4z}{27}}\right)\right)^{1/3} + \left(1 + \frac{27}{2z}\left(1 + \sqrt{1 + \frac{4z}{27}}\right)\right)^{-1/3}\right] \tag{9}$$

for any positive fugacity.

The loosest packing are obtained in the small $z$ limit. Expanding the modulus $\Lambda$ around $z = 0$ and using identity (4), we find

$$\Lambda(z) \underset{z \to 0}{\simeq} z^{2/3} + \frac{z}{3} + \ldots \quad \Longrightarrow \quad \begin{cases} \eta(z) \simeq \frac{2}{3} + \frac{1}{9} z^{1/3} + \ldots \\ \\ S(z) \simeq -\frac{1}{27} z^{1/3} \ln z + \ldots \simeq -\frac{1}{3}(\eta - \frac{2}{3}) \ln(\eta - \frac{2}{3}) + \ldots \end{cases}$$
(10)

The densest packings are found in the large $z$ limit. The asymptotic expression of the modulus reads

$$\Lambda(z) \underset{z \to \infty}{\simeq} z + 1 + \ldots \quad \Longrightarrow \quad \begin{cases} \eta(z) \simeq 1 - \frac{1}{z} + \ldots \\ \\ S(z) \simeq \frac{1}{3z} \ln z + \ldots \simeq -\frac{1}{3}(1 - \eta) \ln(1 - \eta) + \ldots \end{cases}$$
(11)

The entropy curve $S$ as a function of the packing fraction $\eta$ is displayed Fig.3. Stable packing only exist for packing fractions ranging from $\eta = 2/3$ to $\eta = 1$ which represent respectively the loose and dense packings. The entropy $S$ vanishes in these two limits and continuously varies in between, exhibiting a maximum $S = 0.1274$ at $\eta = 0.8057$ for $z = 1$. This packing fraction can be understood as the most probable one : any unbiased dynamics exploring all possible configurations will necessarily leads to this volume fraction.

In this simple case $m = 3$, the results can be easily interpreted in terms of the geometrical structure of the packings. The dense packings $\eta = 1$ correspond to the complete paving of the plane (Fig.4a). The loose packing is also given by a perfectly periodic pattern represented in Fig.4b with a volume fraction equal to 2/3. Between these two limits, packings can be built by filling the free sites of the loose packing, which leads to intermediate volume fractions. However, the reader can easily convince himself (herself) that all stable configurations are not obtained in this way: a simple combinatory calculation shows that the entropy would be in this case equal to

$$S(\eta) = -\frac{1}{9} \ln 3 - \frac{1}{3}(\eta - \frac{2}{3}) \ln(\eta - \frac{2}{3}) - \frac{1}{3}(1 - \eta) \ln(1 - \eta) \quad , \tag{12}$$

which gives the correct asymptotic behaviors. However, the maximum of (12) equals $S \simeq$

0.077 for $\eta = 0.833$ and does not coincide with the transfer matrix result given above. The discrepancy means that there exist others stable packings which do not derive from the loose packing family. One is displayed on Fig.4c. This packing and the related ones obtained by filling more or less the free sites represent another family of stable packings. Both types of arrangements drawn on Fig.4b and Fig.4c can indeed coexist in the same stable configuration. Notice that this geometrical interpretation indicates that the whole entropy curve is indeed physically meaningful, from the loose to the close packings densities. Therefore, the left part of the curve, which corresponds to a negative slope cannot be discarded. In other words, the compactivity $X$ is not constrained to be positive. It is indeed well-known that the entropy of systems with a finite number of configurations (e.g. spins models) is not necessarily an increasing function of the energy. We believe that the fugacity $z$, and equivalently the compactivity $X$, have no particular physical relevance since they can not be controlled in experiments, contrary to the density or the external stress for instance. In our approach, the fugacity is introduced for mathematical purpose only.

This simple example $m = 3$ shows that, even in the case of narrow systems, the statistics of packing is non trivial, exhibiting a whole range of available packing fractions and a maximum of the entropy. However, all the stable arrangements we found exhibit an underlying periodicity that is essentially due to the monodispersed character of the mixture. The periodic boundary conditions imposed in our model thus play an important role, and give rise to problems in the definition of the thermodynamics limit.

We have already mentioned that in the case $m = 3$ the characteristic polynomial (8) of the transfer matrix is a function of $\lambda^3$ only where $\lambda$ is the eigenvalue. Consequently, the generating function (2), according to formula (7), is equal to zero if $L = 3p+1$ or $L = 3p+2$ where $p$ is an integer, and is non zero only for $L = 3p$. From a physical point of view, the degeneracy can be seen as a problem of compatibility between the periodicity imposed by the stability rule (Fig.4), and the periodic conditions imposed to the lattice.

This effect persists when increasing the lattice width $m$, as shown in Table 1. This table presents the degree of degeneracy $q$ of the maximal eigenvalue of the transfer matrix, the maximum entropy and its corresponding packing fraction for increasing values of $m$, in the limit of infinite number $L$ of ranks. The degree of degeneracy $q$ varies in a non trivial way with $m$. We have found numerically that the arguments of the maximal eigenvalues, see (6) and (7), are equally distributed, $\theta_j = 2\pi j/q$, meaning that the number of packings vanishes when $L$ is not a multiple of $q$. Notice that this degeneracy is not usually encountered in classical transfer matrix problems as the elements of the matrix cannot vanish [1] (except for zero temperature).

Since, at fixed $m$, the limit $L \to \infty$ is ill–defined, the thermodynamics limit does not exist from a rigorous mathematical standpoint. However, the physical quantities converge towards well–defined limits (depending on $m$) when $L$ is sent to infinity as a multiple of the degree of degeneracy $q(m)$. One may then wonder whether all these ($m$-dependent) limits converge towards a unique value when $m \to \infty$. Table 1 suggests that it might not be so but no definitive conclusion can be drawn from such rather small systems. It is indeed experimentally well–known that the properties of bidimensional monodispersed packings are dramatically influenced by the boundaries [30]. One way to prevent such peculiar situation consists in mixing two different sizes of particles.

## V. RESULTS FOR POLYDISPERSED PACKINGS

Hereafter, we consider the binary mixture model consisting in small particles that are hexagons, and big particles that are elongated hexagons. In order to compare with the monodispersed case, we first study the simple case of lattice width $m = 3$. The maximal

---

[1] An alternative way to understand the degeneracy is to consider the correlation length of the system which is usually given by the ratio of the second largest eigenvalue (in term of modulus) of the transfer matrix over the first one. However in our degenerated problem there is no second and first eigenvalue as they both have the same modulus and thus the correlation length diverges : there exists a long range order in the stable packings.

eigenvalue $\lambda$ of $\mathcal{Q}_3(z_s, z_b)$ is the largest root of the polynomial

$$P(\lambda, z_s, z_b) = \lambda^5 - z_s\lambda^4 - 3z_b\lambda^3 + (3z_s z_b - z_s^2)\lambda^2 - 6z_s^2 z_b \quad . \tag{13}$$

It is straightforward to check that the expression (9) of the monodispersed case is recovered when the big particle fugacity $z_b$ vanishes. On the opposite, when considering big hexagons only ($z_s = 0$), one finds $\lambda = \pm\sqrt{3z_b}$. The whole entropy curve shrinks to a single point $\eta_b = \frac{5}{6}$ and $S = \frac{1}{6}\ln 3$. This result may be easily understood : when $m = 3$, half of the horizontal lines of the lattice have to be inoccupied and there are three different ways of placing the big particles on every remaining rank.

When mixing big and small particles, the situation becomes more interesting. The polynomial (13) has a non degenerate and real maximal root as soon as $z_b$ is non zero, that is as soon as there is a finite packing fraction of big particles. The introduction of big particle is therefore sufficient to break the periodicity of the stable configurations. This result holds for larger value of $m$ (Fig.2). The role of the boundary conditions being no longer crucial for the bulk properties of disordered packings, the thermodynamics limit is now well defined as suggested by the result of Table 2. Note that instead of considering the entropy per site $S$ as a function of $\eta_s$ and $\eta_b$, we hereafter choose the more common description in terms of $\eta_{tot} = \eta_s + \eta_b$ and of the concentration of big particles $C_b = \eta_b/\eta_{tot}$ which is usually introduced in binary mixture problems. Table 2 shows that the maximum entropy (found for $z_s = z_b = 1$) and its corresponding total packing fraction and concentration of big particles seem to converge to the well–defined limits $S \simeq 0.20$, $\eta_{tot} \simeq 0.82$, $C_b \simeq 0.5$ for increasing $m$.

We then consider the binary mixture on a lattice that is ten sites wide. The size of the transfer matrix $\mathcal{Q}_{10}$ equals $8294 \times 8294$ and thus allows the numerical calculation of the maximum eigenvalue for any fugacities $z_s, z_b$ in a reasonable time [31]. This calculation is carried out for all positive $z_s$ and $z_b$ to obtain the whole surface $S(\eta_{tot}, C_b)$. Notice that the mapping between the $(z_s, z_b)$ plane and the $(\eta_{tot}, C_b)$ plane is highly non trivial. Consequently, it is difficult to scan in a homogeneous way the latter.

Fig.5 displays the two-dimensional contour plot of the entropy $S$ in the $(\eta_{tot}, C_b)$ plane.

We first see that the entropy is non zero, i.e. that stable packings are exponentially numerous, only in a certain range of the total packing fraction and of the concentration of big particles (shaded region on Fig.5). There exist two curves $\eta_{loose}(C_b)$, including points B,C,D and $\eta_{close}(C_b)$, including points A,F,E that delimit the domain of existence of the stable arrangements in the plane $(\eta_{tot}, C_b)$. The calculation of the entropy thus leads to the prediction of the available range of packing fractions for the stable packings.

The close (respectively, loose) packing curve may be in turn divided into two different parts. On the FE (resp. CD) segment, the total packing fraction is constant $\eta_{tot} = 1$ (resp. $\eta_{tot} = \frac{2}{3}$) and the entropy vanishes. On the opposite, from points A to F (resp. B to C), i.e. for $C_b \leq \frac{1}{4}$, the entropy is non zero as can be easily seen in Fig.6 which shows a tree-dimensional view of the entropy surface as a function of the total packing fraction and of the concentration of big particles. Therefore, keeping the fraction of big particles constant $C_b < \frac{1}{4}$ and varying the total volume fraction lead to a curve for the entropy, that is similar to the one obtained in the monosized model of Fig.3, except that the two extreme points corresponding to the loose and dense packings present a non zero entropy. These discontinuities in the entropy curve mean that the number of configurations corresponding to the loose and dense volume fractions grows exponentially with the size of the system. The jump in the entropy curve decreases to zero for fractions of big particle going to zero, that is when the mixture becomes pure, or to one fourth as we shall explained below. Notice that the same kind of entropy discontinuity has been previously observed by Roux et al. in a different model of packing based on links percolation [28].

The above remarks on the general structure of the entropy surface can be better understood by means of a more thorough inspection of the microscopic nature of the packings corresponding to points A to G. The geometrical picture of the packing corresponding to point F is drawn Fig.7a. A simple computation shows that the corresponding concentration $C_b$ indeed equals one fourth. Similarly, point C may be obtained by inserting a column of big particles into the loose packing patterns for small particles displayed in Fig.4b. The resulting packing is given Fig.7b with a volume fraction equal to $\eta_{tot} = \frac{2}{3}$. Points E and D denote the closest and loosest packings made of big particles only, and their geometrical structure may be understood as simple dilatations along the horizontal axis of Fig.4a and 4b respectively. On the bottom side of Fig.5, for $C_b = 0$, one recovers the small particles arrangements extending between B (loosest packing) and A (closest packing). In between, the entropy curve as a function of the volume fraction exhibit the usual bell shape visible on the front of Fig.6. To end with this description, we consider point G lying on the close packing segment EF, at a given concentration $C_b$ ranging between one fourth and unity. The corresponding arrangement is shown Fig.7c. It consists in mixing E and F patterns by first achieving the correct $C_b$. Clearly, to ensure that the resulting void fraction vanishes in the thermodynamical limit, i.e. $\eta_{tot} = 1$, the number of interfaces between E and F patterns has to remain finite, giving rise to a non extensive entropy. Note that a similar argument, based on combining D and C packings can be made to explain the loose packing segment DC.

Far from the borders, the entropy surface roughly looks like a bump whose maximal height, $S_{max} \simeq 0.20$, is much higher than maximal entropies obtained for monodispersed packings. However, one must keep in mind that part of $S_{max}$ is due to mixing effects. While monodispersed arrangements are made of indistinguishable particles, bidispersed mixtures deals with two physically different types of beads. We then must take into account a mixing entropy contribution, which one can evaluate as $\rho_{tot}[-(1 - C_b) \ln(1 - C_b) - C_b \ln C_b]$ where $\rho_{tot}$ equals the total number of particles per site. Maximum of this additional contribution is reached when $C_b$ lies around one half, precisely the result we have found in Table 2. Moreover, its numerical estimate is of the order of $S_{max}$. It might therefore well happen that the entropy excess in bidispersed packings with respect to the monodispersed case mainly results from such mixing effects.

Of higher relevance is the difference between the packings structures taking place on the frontier of the entropy surface (Fig.5) and the central region. As mentioned above, in the former region, packings are expected to strongly reflect the lattice periodicity (Fig.7), as in the monodisperse case. In the latter, stable arrangements will more likely to be disordered, see Fig.2, and less influenced by the boundaries and the lattice parameters, namely its width $m$. This provides explanation for the fast convergence of typical quantities related to the maximum of the entropy surface we have already noted from Table 2.

## VI. CONCLUSION

In this paper, we have introduced a toy model of particles packings, which has allowed us to study the relevance of the concept of packing entropy. First of all, we have shown that the parameters plane (density, relative concentration of big particles) can be divided into two different regions : a first region where stable packings exist and a second region where stable packings cannot be built. The frontier between both regions correspond to the loosest and closest packings.

Secondly, when stable packings exist, we have shown that a huge number of arrangements can be built having the same density and relative concentration of big particles. More precisely, despite the strong constraints resulting from the stability conditions, the number of packings is found to grow exponentially with the area of the lattice and the entropy defined as the logarithm of this number is extensive.

Of course, the hexagon packings model we have presented here is a crude approximation of real packings. The first simplifications lies in the ad hoc stability rule adopted to discriminate between the stable and unstable configurations, which does not take into account the forces network. A second criticism one could address is related to the lattice nature of the model. Understanding to what extent our results could be transposed to packings embedded in a continuous space would be of interest.

**Acknowlegdements** : We thank W. Krauth, L. Tuckerman for useful and stimulating remarks and discussions.

# REFERENCES

* Email: monasson@physique.ens.fr ; preprint 96/36

LPTENS is a unité propre du CNRS, associée à l'Ecole Normale Supérieure et à l'Université de Paris–Sud.

† Email: olivier@ladhyx.polytechnique.fr

LadHyX is a unité mixte de recherches du CNRS, associée à l'Ecole Polytechnique.

TABLES

| width m | degeneracy q | packing fraction $\eta$ | entropy S |
|---|---|---|---|
| 3 | 3 | 0.8057 | 0.1274 |
| 4 | 2 | 0.7500 | 0.0866 |
| 5 | 5 | 0.8006 | 0.0856 |
| 6 | 3 | 0.8323 | 0.1123 |
| 7 | 7 | 0.7907 | 0.0637 |
| 8 | 8 | 0.8219 | 0.0802 |
| 9 | 3 | 0.8368 | 0.1112 |
| 10 | 5 | 0.8052 | 0.0671 |

TABLE I. Degeneracy, packing fraction and maximum entropy for different widths of the lattice in the monodispersed case ($z = 1$). Remark that the entropy does not seem to converge toward a unique limit as $m$ grows, compare to Table 2.

| width m | packing fraction $\eta_{tot}$ | concentration $C_b$ | entropy S |
|---|---|---|---|
| 3 | 0.7891 | 0.5608 | 0.2634 |
| 4 | 0.8050 | 0.5542 | 0.2212 |
| 5 | 0.8105 | 0.5523 | 0.1894 |
| 6 | 0.8066 | 0.4742 | 0.1977 |
| 7 | 0.8169 | 0.5116 | 0.2007 |
| 8 | 0.8197 | 0.5522 | 0.1968 |
| 9 | 0.8177 | 0.5151 | 0.1941 |
| 10 | 0.8185 | 0.5041 | 0.1959 |

TABLE II. Total packing fraction, concentration of big particles and maximum entropy for different widths of the lattice in the polydispersed case ($z_s = z_b = 1$). Remark that packing fractions, relative concentrations of big particles and entropies seem to converge toward well–defined limits as $m$ grows, compare to Table 1.

# Figure Captions

Figure 1: The stability condition adopted in our packing hexagon model; a) hexagon A is stable; b) hexagon A is unstable along the arrow direction.

Figure 2: Example of a stable packing in the bidispersed model for $m = 10$.

Figure 3: Packing entropy $S$ versus packing fraction $\eta$ for the monodispersed case m=3.

Figure 4: Monodispersed packings for m=3: a) close packing ($\eta = 1$); b) loose packing ($\eta = 2/3$) c) packing with $\eta = 3/4$. The underlying periodic cells are drawn; their heights are multiple of three (see text).

Figure 5: Contour plot of the entropy $S$ for the bidispersed model. Stable packings only exist in the shaded region of the plane ($\eta_{tot}$,$C_b$).

Figure 6: Three–dimensional plot of the entropy $S$ versus packing fraction $\eta_{tot}$ and concentration of big particles $C_b$.

Figure 7: a) Packing at point F; b) packing at point C; c) packing at point G. See Figures 5 and 6 for the definition of the points.

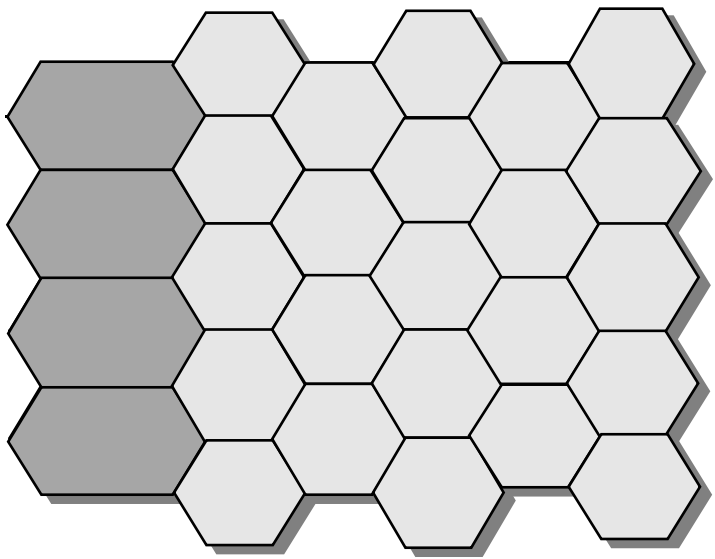

Fig 7a

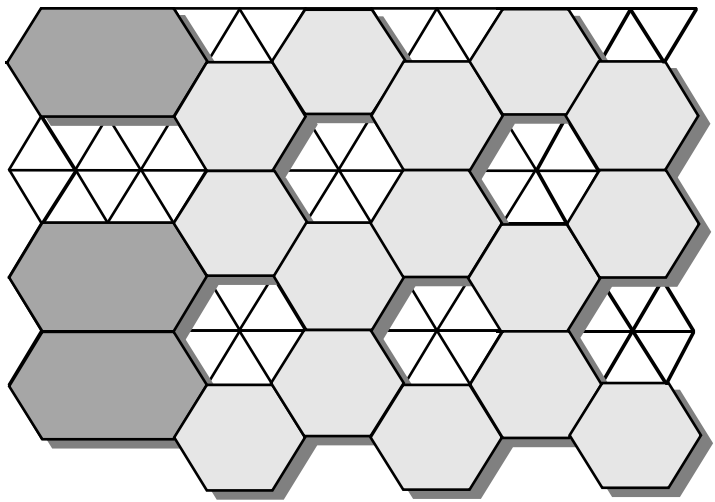

Fig 7b

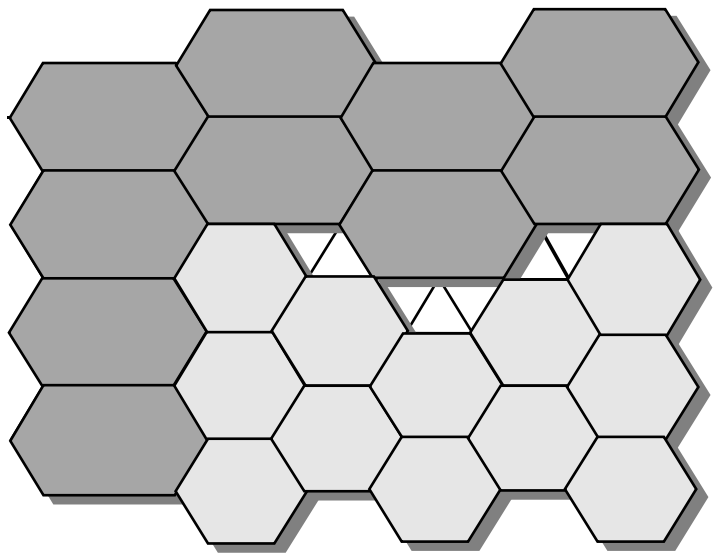

Fig 7c